\documentclass[twocolumn,titlepage]{revtex4-1}
\usepackage[english]{babel}
\usepackage{setspace} 
\usepackage{amsmath}
\usepackage{multirow}
\usepackage{graphicx}
\begin{document}
\title{Theoretical analysis of thermophoretic experimental data}

\author{J.M. Sancho}

\affiliation{Universitat de Barcelona, Departament d F\'isica de  la Mat\`eria Condensada,\\
Universitat de Barcelona Institute of Complex System (UBICS)\\
Mart\'i i Franqu\'es, 1. E-08028 Barcelona. Spain}


\begin{abstract}
 
Thermophoresis is a transport phenomenon induced by a temperature gradient.
Very small  objects dispersed in a fluid medium and in a temperature gradient present a non homogeneous steady density. Analysing this phenomenon within the theoretical  scenario of non interacting Brownian motion one can assume that those particles are driven by a spatially dependent mechanical  force. This implies the existence of a potential which was derived in a previous work. From this potential the qualitative properties of the force and the Soret coefficient were obtained. Nevertheless a quantitative correlation between the theory and the experimental data were not consistently proved. Here it is presented a methodology to match this theory with the experimental data, which it  is used to analyse  the experimental information of sodium docecyl sulfate (SDS) micelles.  

\end{abstract}

\date{\today}
\maketitle
\noindent{\it Keywords\/}: Thermophoresis, Soret coefficient.

\section{Introduction}
\label{intro}

It is a well established transport phenomenon  that small particles in a fluid medium and under a temperature gradient present a biased motion either toward the hotter side (thermophilic) or toward the colder one (thermophobic), staying around  a preferred temperature $T^*$ where the particle steady density $c_{st}$ presents a maximum. This effect is generic for a large variety of small particles such as: plastic micro spheres, micelles, DNA and RNA  polymers, etc. There are a lot of experimental data on the behaviour of these particles under different conditions in a temperature range $\sim (275,310)$K \cite{Wurger}.

 In a typical situation, a single particle or an ensemble of colloids  are confined within a rectangular plate region filled with some fluid and prepared under a temperature gradient. In the steady state an inhomogeneous density $c_{st}(x)$ or probability distribution $P_{st}(x)$ is achieved. Most of the experimental situations can be considered as one dimensional systems. 

This  temperature dependent transport phenomenon is measured by  the Soret coefficient $S_T$, which is defined as \cite{Iacopini06},

\begin{equation}
 \frac{d c_{st}}{dx} = - c_{st} S_T \frac{dT}{dx}.
 \label{Soret1}
\end{equation}

As the temperature gradient can be controlled externally $T=T(x)$, this probability can be expressed in terms of the variable $T$ instead of $x$.
\begin{equation}
 S_T(T)= - \frac{d}{dT} \ln c_{st}(T)=- \frac{d}{dT}\ln P_{st}(T)
 \label{Soret}
\end{equation}

In the case of very dilute concentration of particles, $c_{st}(x)$ can be identified with the probability density distribution $P_{st}(x)$ of a single particle. Given this observable, the Soret coefficient is evaluated at different points and plotted versus the temperature at these points. 
Although other types of information can be extracted  \cite{Wurger},  this work will use only the above information.

 In these plotts the role of other important experimental variables are observed:

\begin{itemize}
\item The particle size measured either by the radius of polystyrene beads \cite{Duhr06,Braibanti08}, the molecular weight \cite{Iacopini06}, or the number of bases of RNA and DNA strands \cite{Reichl14}.

\item The surface chemical preparation of the Brownian particle \cite{Braibanti08}.

\item The  chemical substances dissolved in the fluid medium: salt, pH, etc, \cite{Iacopini03}.

\end{itemize}

The first element of the theoretical approach starts by assuming that the steady density $c_{st}(x)$ has been originated by the overdamped Brownian motion of non interacting particles under a thermophoretic force $F(x)$ derived by a potential $V(x)=k_B {\hat V}(x)$. 
In this framework one can write a dynamical Fokker-Planck equation for the probability of a single particle $P(x,t)$ in a temperature gradient \cite{VanKampen} and in appropriate dimensions, it is

  \begin{equation}
 \frac{\partial P(x;t)}{\partial t}=  \frac{\partial}{\partial x} \frac{1}{ \gamma(x)} \left[ {\hat V}'(x) +
 \frac{\partial }{\partial x}  T(x)  \right]P(x;t),
\label{FP3}
\end{equation} 
where the nonequilibrium steady probability is,
  \begin{equation}
P_{st}(x) = \frac{N}{ T(x)} \exp\left( -\int^x \frac{{\hat V}'(x')}{T(x')} d x' \right). 
\label{overP}
\end{equation}

From this result the Soret coefficient (\ref{Soret}) can be calculated, 

\begin{equation}
 S_T(T)= \frac{1}{T} + \frac{1}{T} \frac{d {\hat V}(T)}{d T}.
 \label{ST-eq}
\end{equation}

The first contribution represents an effect common to all Brownian particles.
Although this effect is small, it cannot be avoided in most of the experiments because $S_T(x)$ is also very small. Moreover as we are interested in the knowledge of the potential $V(x)$ it is mandatory
  to subtract  the contribution $1/T(x)$ from the experimental data. Accordingly we define the shifted Soret coefficient $S_F (T)$,
 \begin{equation}
S_F(T)= S_T(T)- \frac{1}{T} =  \frac{1}{T}\frac{d {\hat V}(T)}{d T}.
\label{maineq}
\end{equation}
This equation connects the empirical data of $S_T(T)$ with the potential ${\hat V}(T)$.

Extending this result we can also obtain an expression for the thermophoretic force \cite{Helden15,Pedersen},
\begin{equation}
 F_T(T) = - \frac{d V(T(x))}{dx}= - k_B  \frac{d {\hat V}(T)}{dT}\frac{dT}{dx}=
 -k_BTS_F \frac{dT}{dx}, 
 \end{equation}

Even this force is a very small quantity (fN), it has been measured in some experiments \cite{Helden15,Pedersen}.

The second element of the theoretical approach was the derivation of a thermophoretic potential \cite{Sancho18}, using statistical physics,
  \begin{equation}
 V(x(T))= k_B {\hat V}(T)= k_B N_0 T \, \ln \left( 1 + \chi e^{T_a/T} \right).
  \label{theoeq}
  \end{equation} 

From this analytical expression one can obtain an explicit formula for the shifted Soret coefficient (\ref{maineq}),
\begin{equation}
S_F(T)=\frac{N_0}{T} \left[\ln \left( 1 + \chi e^{T_a/T} \right) - \frac{T_a}{T} \frac{\chi e^{T_a/T}}{1 + \chi e^{T_a/T}} \right], 
\label{STtheoric}
\end{equation}
where it depends on three physical  parameters:

\begin{itemize}
\item $N_0$ is the number of  absorbing sites per colloid. This is the extensive parameter of the theoretical approach.
\item $\chi \, \in \, [0,1]$, is a measure of the  volume fraction of the molecules dissolved in the fluid which can be attached to the colloid.
\item $T_a= \varepsilon_a /k_B$, where $\varepsilon_a$ is the binding energy of the dissolved molecules attached to the particle surface.
  
\end{itemize}    

The aim of this work is to establish a well funded theoretical procedure to match the theory with the experiments obtaining the physical parameters $(N_0, T_a, \chi)$ from the experimental data.
 
In the previous work \cite{Sancho18} the analysis of the experimental data in \cite{Braibanti08} was performed by a nonlinear fit of the theoretical expression of $S_T(T)$. This fit was non conclusive and it will be revisited in the next section, where it is presented the analytical methodology to match the theory and experimental data, discussing the different role of each theoretical parameter. As a test of the approach, in the Section 3 it is presented the study of a particular experimental set-up where all the theoretical parameters are obtained from the experimental data. Finally the conclusions  summarise the main results of this work.

\section{The connection between  experimental data and theory}

Let us start with  an analysis of the experimental information available from Refs.\cite{Iacopini06,Duhr06,Braibanti08,Iacopini03,Helden15,Piazza04}. 
Several generic observations can be extracted from experimental data of $S_T(T)$ versus:

i) In all the experiments it can be calculated a $T^*$ where $S_T(T^*))=0$.

ii) The behaviour of $S_T(T)$ is quasi linear in $(T-T^*)$ with a rate increasing with the particle size \cite{Iacopini06,Braibanti08,Reichl14}.

iii) Given the observed curvature near $T^*$, the quadratic is negative.

iv) In some experiments $S_T(T)$ has a maximum for larger $T$ or it saturates \cite{Reichl14}, with data more dispersed.

It was shown  \cite{Sancho18} that the potential $V(x)$ fulfils the first three observations, but no systematic quantitative study of its analytical properties was given at that time. Moreover it is clear from this behaviour that the shifted Sored coefficient $S_F(T)$ fulfils the same analytical properties
with a small translation of $T^*$.

From this information it can be proposed the following empirical expansion of $S_F(T)$ to match this information with the theory,

\begin{equation}
 S_F(T) = a (T-T^+) - \frac{b}{2} (T-T^+)^2 +  \cdots,
\label{polinomio}
\end{equation}
assuming  that the theory will apply very close to  $T^+$.
This form defines  three empirical parameters $a, b, T^+$ to be extracted from the experimental data by a simple polynomial fit.
As we will see in the forthcoming figures, a plot of the theoretical $S_F(T)$ versus $T$ for different theoretical parameters presents the same empirical behaviour near $T^+$ but with a very small curvature in all cases. The conclusion is that the experimental parameter $b$ is not well described by the theory because according to the experiments it dominates for larger values of $(T-T^+)$ where other theoretical or experimental factors, not incorporated in the theory, are relevant \cite{Reichl14}.

Then the useful experimental parameters are $(a,T^+)$, which can be obtained from the theory by using the definitions 
\begin{equation}
 S_F(T^+)=0, \qquad  a= \frac{d S_F(T)}{dT} \Big|_{T^+}.
 \label{T*-a}
\end{equation}

The next step is to relate these parameters with the theoretical ones $(N_0, T_a, \chi)$. This is not a straightforward task but one can assume, as it will be tested below,  that the parameter $\chi$ is  small enough  to expand the equation (\ref{STtheoric}) up to second order in  $\chi$,
\begin{eqnarray}
 S_F(T)=\frac{N_0\, \chi\, e^{T_a/T}}{T}  \times
\nonumber\\
\left[ \left( 1- \frac{T_a}{T}\right)-
\frac{\chi\,e^{T_a/T}}{2}(1 - 2\frac{T_a}{T}) + \cdots \right].
\label{SF}
\end{eqnarray}
Now after some quite long algebra one can get the following analytical results,
\begin{equation}
T_a \simeq T^+(1+ \frac{e}{2} \chi),   \qquad  N_0 \chi\simeq \frac{ a T^{+2}}{e}.
\label{Ta-N0}
 \end{equation}
 
These equations are the main relevant theoretical results that will be used to match  the experimental information. Nevertheless,  as in a set of experimental data we have to get three theoretical parameters with only two equations, we will need another different set of data with two common parameters with the previous set. Now using these two sets we have four parameters and four equations, but the procedure needs further analysis.

\begin{figure}
\centering
\vskip-5mm
\includegraphics[scale=0.3,angle=-90]{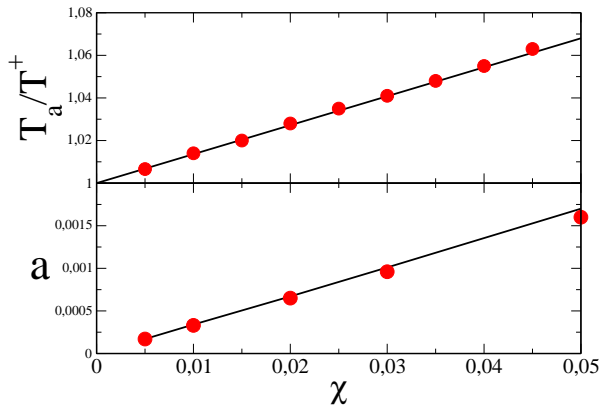}
\caption{Top: $T_a/T^+$ versus $\chi$ with $T_a=290$ K. Bottom: Parameter $a$ versus $\chi$ with $T^+=285 $K. $N_0=1000$. Full lines correspond to the analytical approximation (\ref{Ta-N0}), and dots are obtained numerically from the exact Eq. (\ref{STtheoric}).
}
\label{Ta-chi}
\end{figure}

First we have to test the approximations in (\ref{Ta-N0}). Figs. \ref{Ta-chi}  shows that  the  expresions (\ref{Ta-N0})  are good enough for $\chi \le 0.05$. 

 Next we proceed with the analysis of the shifted Soret coefficient dependence on each of the parameters $N_0, T_a, \chi$  from  the theory (\ref{STtheoric}). Three different procedures are explored. In each procedure two series of experiments with two common  parameters are analysed. The aim is to determine which is the best experimental procedure  to extract the theoretical parameters.


 {\bf Parameter $N_0$.} In Fig. {\ref{EPJE-3} it is shown that the role of the extensive quantity $N_0$ is to increase the parameter $a$ when the other two parameters are kept constant. The behaviour can be compared with Fig. 3 of Ref. \cite{Iacopini06} for some polyelectrolytes (NaPSS) with three different molecular weights.
Other experimental examples are:  Spherical micelles (NaPSS) shown in  Fig-3 of Ref. \cite{Iacopini06}, colloids (polystyrene latex) in  Fig-1 of Ref. \cite{Braibanti08} and ssDNA molecules with different number of bases in Fig-3c of Ref. \cite{Reichl14}. Nevertheless the experimental information is not enough to determine the values of the theoretical parameters from Eqs (\ref{Ta-N0}). As the parameters  $T_a$ and $\chi$  are fixed we have only an equation for $(T_a,\chi)$ and then only  three independent equations to get four parameters. We  can  determine only the factor $N_{01}/N_{02} = a_1/a_2$ from a pair of two experiments with two different sizes. This is the reason why the fit in \cite{Sancho18} was not consistent. In fact neither the parameter $T_a$  nor the parameter $\chi$ were  the same for the three sets of data analysed.

\begin{figure}
\vskip3mm
\centering
\includegraphics[scale=0.3,angle=0]{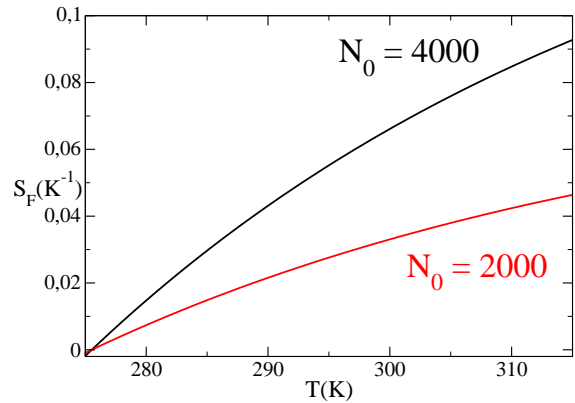}
\caption{Shifted Soret coefficient versus $T$ for two colloid sizes $N_0$. Fixed parameters are $T_a=285$K, and  $\chi = 0.025$. 
}
\label{EPJE-3}
\end{figure}

{\bf Parameter  $T_a$.}
In this case the parameters $N_0, \chi$ are fixed.  One can prove from Eqs. (\ref{Ta-N0}) that we get only two expresions,
\begin{equation}
\frac{T_{a1}}{T_{a2}} =\frac{T_1^+}{T_2^+} \qquad  \frac{a_1}{a_2}=\frac{T_1^{+2}}{T_2^{+2}}.
\end{equation}
The first one gives a linear relation between $T_{a1}$ and $T_{a2}$, and the second equation does not includes any theoretical parameter. It is only a relation between empirical data.
This situation cannot be used to get the absolute value of the theoretical parameters.
Similar behaviour appears in some experiments of lysozyme solutions, see Figs. 2, 3 and 4 in Ref.
\cite{Piazza04}. 

The next case will show one  experimental procedure that will permit  to make effective the connection with the theory. 

\begin{figure}
\vskip5mm
\centering
\includegraphics[scale=0.3,angle=-90]{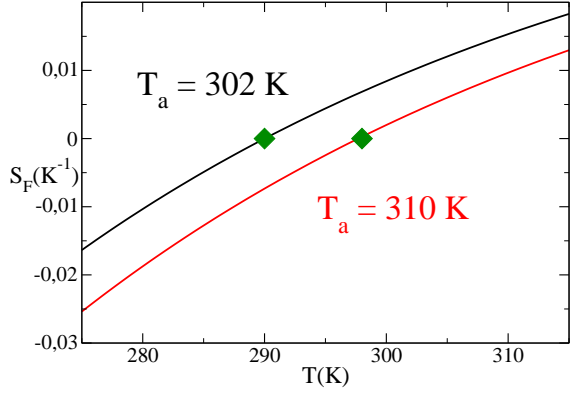}
\caption{Shifted Soret coefficient versus $T$ for two values of $T_a$. Green points are the corresponding temperatures $T^+$. Fixed parameters $N_0 = 1000$ and  $\chi = 0.03$.
}
\label{PCCP}
\end{figure}

{\bf Parameter $\chi$.}
In Fig.\ref{EPJE-2} it is explored the role of the parameter $\chi$ for two cases  with $(N_0,T_a)$ fixed. It is seen  that the experimental parameters $T^+$ and $a$ are well differentiated in these two different sets (1,2).


\begin{figure}
\vskip5mm
\centering
\includegraphics[scale=0.3,angle=-90]{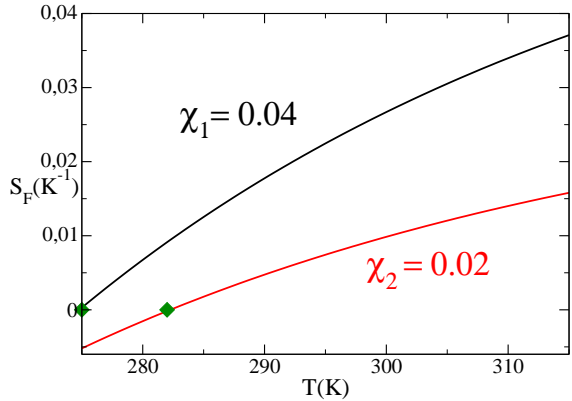}
\caption{Shifted Soret coefficient versus $T$ for two values of $\chi$. Green points are the corresponding temperatures $T^+$. Fixed parameters $T_a=290$K, and  $N_0 = 1000$. 
}
\label{EPJE-2}
\end{figure}

Evaluating the parameters ($a_1, T_1^+$) and ($a_2, T_2^+$) by the fit  (\ref{polinomio}) of the experimenta data, we can calculate the auxiliary quantities 
\begin{equation}
 a= \frac{a_1}{a_2} , \qquad d=\frac{T_1^+}{T_2^+}.
\label{a}
 \end{equation}
Then, using Eqs. (\ref{Ta-N0}) for these two cases, it is obtained  the following two linear equations,
\begin{equation}
 d(1+\frac{e}{2} \chi_1)= 1+\frac{e}{2} \chi_2, \qquad \qquad \chi_1= a d^2 \chi_2,
 \label{solution}
\end{equation}
with the solution,
\begin{equation}
 \chi_2 = \frac{2}{e} \frac{d-1}{1- a d^3},
\label{chi}
 \end{equation}
if $ a> 1, a d^3 > 1$. Finally the parameter $\chi_1$ is obtained from (\ref{solution}) and  
$N_0$ and $ T_a$ from (\ref{Ta-N0}). 
This is the best  procedure to get the theoretical parameters from experimental data. 

\section{Analysis of an experiment}
An experimental situations which can be studied using the last procedure  is reported  in  Fig. 2 of Ref. \cite{Iacopini06}, where sodium docecyl sulfate (SDS)  micelles are studied for two different concentrations of [NaCl] and two micelles concentrations $c$.
The first step is to subtract the contribution $1/T$ from the experimental data. Then,  the case of lower concentration of micelles $c=10$g/l is selected to get the  parameters. The experimental parameters $a_1,a_2,T_1^+, T_2^+$ are calculated by fitting the polynomial (\ref{polinomio}). 
Then, using Eqs. (\ref{Ta-N0}) and (\ref{chi}) the  theoretical parameters  $(T_a, N_0,\chi_1, \chi_2)$ are also obtained (See the first two lines in Table \ref{table1}). 
 As it is seen in the figure \ref{SDS} the agreement (black symbols) is good for small values of $(T-T^+)$ as expected.  Now we can repeat the procedure for the experimental data of $c=20$g/l, obtaining the results in the last two lines of  Table \ref{table1}.

\begin{figure}
\centering
\includegraphics[scale=0.35,angle=-90]{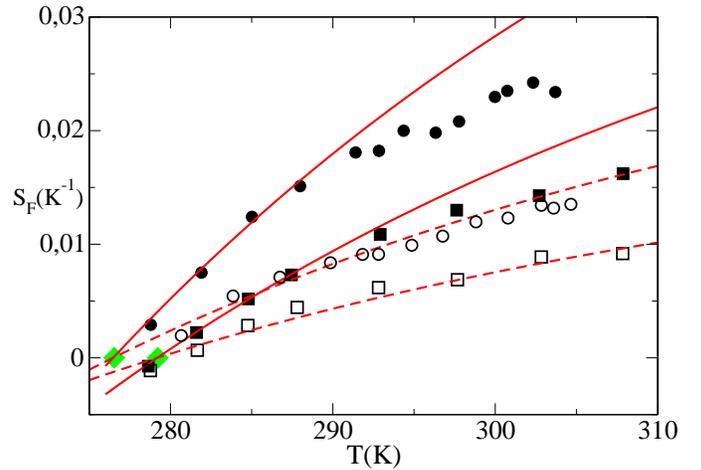}
\caption{Shifted Soret coefficient $S_F(T)$  versus $T$ for two concentrations of salt  NaCl: $20\,$mM (circles) and $10\,$mM (squares), and two micelles concentrations $c=10$g/l (full symbols) and $c=20$g/l (open symbols). Dots are experimental data from Fig. 2 in \cite{Iacopini06}, and  lines are from the equation (\ref{STtheoric}) with the theoretical parameters of Table \ref{table1}. 
}
\label{SDS}
\end{figure}

 \begin{table}[htb]
\begin{center}
\begin{tabular}[c]{|l|l|l|l|l|l|l|l|l|}
\hline
 c(g/l) &[NaCl] & \,\,a(K$^{-2}$) & \,$T^+$(K) && \,\,\,\,\,\,$\chi$ &  $T_a$(K) &\,\,\, $N_0$ \\
 \hline  
 10&10 mM &9.9\,$10^{-4}$&  279.22& &{\bf 1.31\,}$10^{-2}$&{\bf 284.18}& {\bf 2174}\\
10& 20 mM &1.6\,$10^{-3}$& 276.54& &{\bf 2.03\,}$10^{-2}$ &284.18&2174\\
\hline
 20&10 mM &6.2\,$10^{-4}$& 280.28& &1.31\,$10^{-2}$&284.18&{\bf 1000}\\
20& 20 mM &7.5\,$10^{-4}$& 276.83& &2.03\,$10^{-2}$ &284.18&1000\\
\hline
\end{tabular}
\end{center}
\caption{Fitted parameters $ a , T^+$  from the experimental data of Fig.2 in \cite{Iacopini06}, and calculated  microscopic parameters $ T_a, \chi_1, N_{01}, \chi_2, N_{02}$ from Eqs.(\ref{Ta-N0}, \ref{a}, \ref{chi}). The bold faced values are the five physical theoretical parameters.} 
\label{table1}
\end{table}

For the case of large micelles concentration it was assumed that the values of $T_a$ and both $\chi$'s  have to be same as in the low concentration case , but with different $N_0$ because it was assumed that the available surface to bind has been reduced. In these cases the final value used $N_0=1000$  is chosen on order to optimise the fitting procedure.
The agreement between theory and experiments is good given the dispersion of the experimental data. The four set or data (12  parameters)  have been fitted with only five different physical parameters.   
It is worth to mention that the biding energy of salt to the micelles surface is $ \varepsilon_a= K_B T_a = 3.92 \,10^{-21}$J.

 Once  the theoretical parameters are known one can repeat the experiments by changing $N_0$ or the salt concentration to get further  information on the system.

\section{Conclusions}
It has been shown that a theoretical  model based on statistical physics  can be used to extract information  from  a set of thermophoretic experimental data. With this information further experiments can be performed to know how the model parameters will change  on the different experimental preparations. Moreover, herein we take advantage of the later example to propose some modification in the  experimental set-ups to get more easily the theoretical parameters.

For the case of very small particles it is mandatory  to subtract the thermal contribution $1/T$ from the experimental data of $S_T(T)$. 

A very important fact  is to use  particles with a large enough value of $N_0$. This is easily done using  colloidal particles with very large radius as in Refs. \cite{Braibanti08,Helden15}. A single big particle tracking  could be  the perfect procedure to compare with this theory \cite{Helden15}.

 In a given experimental situation it would be necessary to perform a series of  different temperatures close to $T^*$, to  better fix  the parameters $(a, T^+)$.
Then it should follow a second set with a change of another experimental variable such salt concentration chosen  to exhibit appreciable differences on $(a, T^+)$ with respect to the first set. 
 
 After completing this process  the theoretical parameter, mainly $N_0$,  is well controlled, and one can proceed with the study of other situations:  particle surface treatments, different salts, pH, fluid medium,...etc. 
 
 The knowledge of the theoretical parameters could be of utility in other contexts involving temperature gradients. $N_0$ can be related with the surface treatment of a bead, or with the number of monomers in RNA strands,..etc. The parameter $T_a$ could be useful in absorption experiments, and $\chi$ can be related with pH or with the ionic strength of the medium.
 



\end{document}